\pdfoutput=1
\documentclass[sigconf,nonacm]{acmart}
\setcopyright{none}
\renewcommand\footnotetextcopyrightpermission[1]{}
\usepackage{tcolorbox}
\usepackage{multirow}
\usepackage{enumitem}
\AtBeginDocument{%
  }

\newboolean{showcomments}
\setboolean{showcomments}{false}         
\ifthenelse{\boolean{showcomments}}
  {
    \newcommand{\nb}[2]{%
      {\small\bfseries\sffamily\textcolor{green}{#1}}%
      ~{\sf\small$\blacktriangleright$\textit{\textcolor{red}{#2}}$\blacktriangleleft$}%
    }
  }
  {
    \newcommand{\nb}[2]{}
    
  }

\newcommand\gunel[1]{}

\newcommand\jie[1]{\nb{Jie}{#1}} 

\begin{document}

\title{How Does Chunking Affect Retrieval-Augmented Code Completion? A Controlled Empirical Study}


\author{Xinjian Wu}
\affiliation{%
  \institution{King's College London}
  \city{London}
  \country{UK}}
\email{xinjian.wu@kcl.ac.uk}

\author{Jingzhi Gong}
\affiliation{%
  \institution{King's College London}
  \city{London}
  \country{UK}}
\email{jingzhi.gong@kcl.ac.uk}

\author{Gunel Jahangirova}
\affiliation{%
  \institution{King's College London}
  \city{London}
  \country{UK}}
\email{gunel.jahangirova@kcl.ac.uk}

\author{Jie M. Zhang}
\authornote{Corresponding author.}
\affiliation{%
  \institution{King's College London}
  \city{London}
  \country{UK}}
\email{jie.zhang@kcl.ac.uk}

\renewcommand{\shortauthors}{Wu et al.}

\begin{abstract}
Retrieval-augmented generation (RAG) pipelines for code completion rely on chunking to segment source files into retrievable units, yet chunking strategies are typically adopted without empirical justification, and
practitioner recommendations are notably inconsistent.
We present a controlled empirical study isolating the effect of chunking on code completion quality by crossing four representative strategies (Function, Declaration, Sliding Window, and cAST) with four retrievers, five generators, and nine parameter configurations on two benchmarks (RepoEval and CrossCodeEval), totaling 864 experimental settings\jie{the task is missing here: what coding task do we focus on? Please add}.
Our results reveal that chunking strategy has a statistically significant effect on
RAG-based code completion.
Contrary to intuition,
chunking based on functions
underperforms all other strategies by 3.57--5.64 percentage points on RepoEval (Cliff's $\delta = -1.0$), while the remaining chunking strategies perform comparably. Our further analysis demonstrates that this observation holds across all retriever--generator combinations.
We also find that cross-file context length is the dominant parameter: doubling from 2{,}048 to 8{,}192 tokens yields up to 4.2 percentage points of improvement, whereas chunk size has a weaker, non-monotonic effect. On the cost--quality Pareto front, Sliding Window and cAST dominate both benchmarks; Function chunking is never Pareto-optimal.
\end{abstract}





\maketitle

\section{Introduction}
Retrieval-augmented generation (RAG) has become a widely adopted approach for repository-level code completion, where cross-file context is retrieved and prepended to the prompt of a large language model (LLM) to improve completion quality~\cite{taoRetrievalAugmentedCodeGeneration2025, zhangRepoCoderRepositoryLevelCode2023}. A typical RAG pipeline consists of three stages: \emph{chunking} source files into retrievable units, \emph{retrieving} the most relevant chunks given a query context, and \emph{generating} the completion conditioned on both the query and the retrieved context. While prior work has focused on improving the retriever with better embedding models and retrieval algorithms~\cite{wang_rlcoder_2024, jiang_aligncoder_2026} and the generator with larger or fine-tuned code LLMs~\cite{wu_repoformer_2024}, the chunking stage has received little systematic investigation by comparison. The optimal segmentation granularity and whether structure-aware methods outperform \gunel{not sure if we need to use the french way of writing the word "naive"} naive sliding-window approaches remain open questions, as existing work lacks controlled cross-strategy comparison~\cite{zhangCASTEnhancingCode2025}. This gap is consequential: chunking determines the atomic units that the retriever can select, and a poor chunking strategy may limit downstream completion quality even when the retriever and generator are strong.

Existing studies that do address chunking either propose a single new method and evaluate it against a narrow baseline~\cite{zhangCASTEnhancingCode2025}, or adopt a fixed chunking scheme as an unexamined preprocessing step within a broader retrieval or generation framework~\cite{zhangRepoCoderRepositoryLevelCode2023, wang_rlcoder_2024, wu_repoformer_2024}. Practitioner guidance is also \textbf{inconsistent}: Google recommends chunking along natural code boundaries such as functions, classes, or modules\footnote{\url{https://cloud.google.com/blog/products/ai-machine-learning/context-aware-code-generation-rag-and-vertex-ai-codey-apis}}, and Mistral~AI advises splitting by meaningful code units using a syntax-tree parser\footnote{\url{https://docs.mistral.ai/capabilities/embeddings/rag_quickstart}}, yet Codestral Embed defaults to fixed-size sliding-window chunks of 3{,}000 characters with 1{,}000 characters overlap\footnote{\url{https://mistral.ai/news/codestral-embed}}. Even within size-based approaches, LlamaIndex identifies 1{,}024 tokens as optimal for RAG\footnote{\url{https://www.llamaindex.ai/blog/evaluating-the-ideal-chunk-size-for-a-rag-system-using-llamaindex-6207e5d3fec5}}, while LangChain offers language-specific code splitters without recommending a default configuration\footnote{\url{https://docs.langchain.com/oss/python/integrations/splitters}}. These recommendations disagree on both the fundamental approach (structure-aware versus fixed-size) and parameterization (characters versus tokens, chunk size), yet none is supported by controlled evaluation. While some prior work treats chunking as a primary contribution~\cite{zhangCASTEnhancingCode2025} or evaluates chunking strategies across retrievers and parameter configurations without varying the generator~\cite{galimzyanovPracticalCodeRAG2025}, no study examines chunking's effect across the full RAG pipeline, jointly varying retrievers, generators, and parameter settings, to isolate its independent contribution to downstream completion quality. We fill this gap with a controlled empirical study of retrieval-augmented code completion that crosses four chunking methods (Function, Declaration, Sliding Window, and cAST) with four retrievers, four code LLMs and one general-purpose LLM, and nine parameter configurations on two established benchmarks, RepoEval~\cite{zhangRepoCoderRepositoryLevelCode2023} and CrossCodeEval~\cite{ding_crosscodeeval_2023}, totaling 864 experimental settings\jie{again the task is missing here: code completion}.

Our controlled study yields four findings concerning chunking's effect on completion quality, its stability across pipeline configurations, parameter sensitivity, and cost--quality trade-offs. First, chunking strategy has a measurable effect: Function chunking performs worse than all other strategies by 3.57--5.64 percentage points (pp) Exact Match (EM), while Declaration, Sliding Window, and cAST perform comparably ($\leq$2.1\,pp EM difference). Second, this ordering of chunking strategies is stable across all retriever and generator combinations: retriever choice accounts for $\leq$1.11\,pp EM variation on RepoEval, far less than the 3.96\,pp gap between the best and worst chunking strategies. Third, cross-file context length is the dominant parameter: increasing it from 2{,}048 to 8{,}192 tokens yields up to 4.2\,pp EM gain, whereas chunk size has a weaker, non-monotonic effect ($\leq$1.9\,pp). Fourth, Sliding Window and cAST are Pareto-optimal on the cost--quality trade-off on both benchmarks; Function chunking is never Pareto-optimal despite its lower token cost.

\noindent The primary contributions of this paper are:
\begin{enumerate}[leftmargin=*, nosep]
\item The first controlled empirical study that isolates the chunking dimension in retrieval-augmented code completion, with retriever, generator, and parameters held constant to measure chunking's independent effect.
\item A controlled comparison of structure-aware and structure-agnostic chunking strategies across the full RAG pipeline; structure-aware methods do not outperform Sliding Window on quality or cost efficiency.
\item Evidence-based parameter recommendations from ablation across chunk sizes and cross-file context lengths; cross-file context length is the dominant tuning dimension for practitioners.
\item A replication package comprising the full evaluation pipeline across four chunking methods, four retrievers, five generators, and nine parameter configurations on two Python benchmarks (RepoEval and CrossCodeEval), available at \url{https://doi.org/10.5281/zenodo.19228777}.
\end{enumerate}

\section{Background and Related Work}

This section defines the three pipeline stages---chunking, retrieval, and generation---whose interaction our study investigates and positions our work relative to prior retrieval-augmented code completion systems. \gunel{I feel like in this sentence "chunking, retrieval, and generation" eithed needs to be put into brackets or surrounded by dashes for better clarity.}
\subsection{Chunking}
\label{sec:chunking}

Chunking is the process of segmenting a document into smaller retrievable units for downstream retrieval. Given a document $d$, a chunking function $\mathcal{C}$ produces a sequence of fragments:
\begin{equation}
  \mathcal{C}(d) = (c_1, c_2, \dots, c_l)
\end{equation}
where each chunk $c_j$ is a contiguous span of the original document, optionally augmented with metadata (e.g., file path, line range, chunk size, and enclosing node count). Applied across all documents in a corpus, chunking produces the full set of retrievable units $C = \bigcup_{d} \mathcal{C}(d)$ that serves as input to the retrieval stage.

In natural language processing, chunking for RAG pipelines has received growing attention since Lewis et al.~\cite{lewis_retrieval-augmented_nodate} introduced the RAG paradigm. Common strategies include fixed-size splitting, sentence-boundary splitting, semantic chunking by embedding similarity, and hierarchical clustering~\cite{sarthi_raptor_2024}. These studies demonstrate that chunk granularity and boundary alignment affect retrieval precision and downstream quality.

\begin{figure}[!t]
  \centering
  \includegraphics[width=\columnwidth]{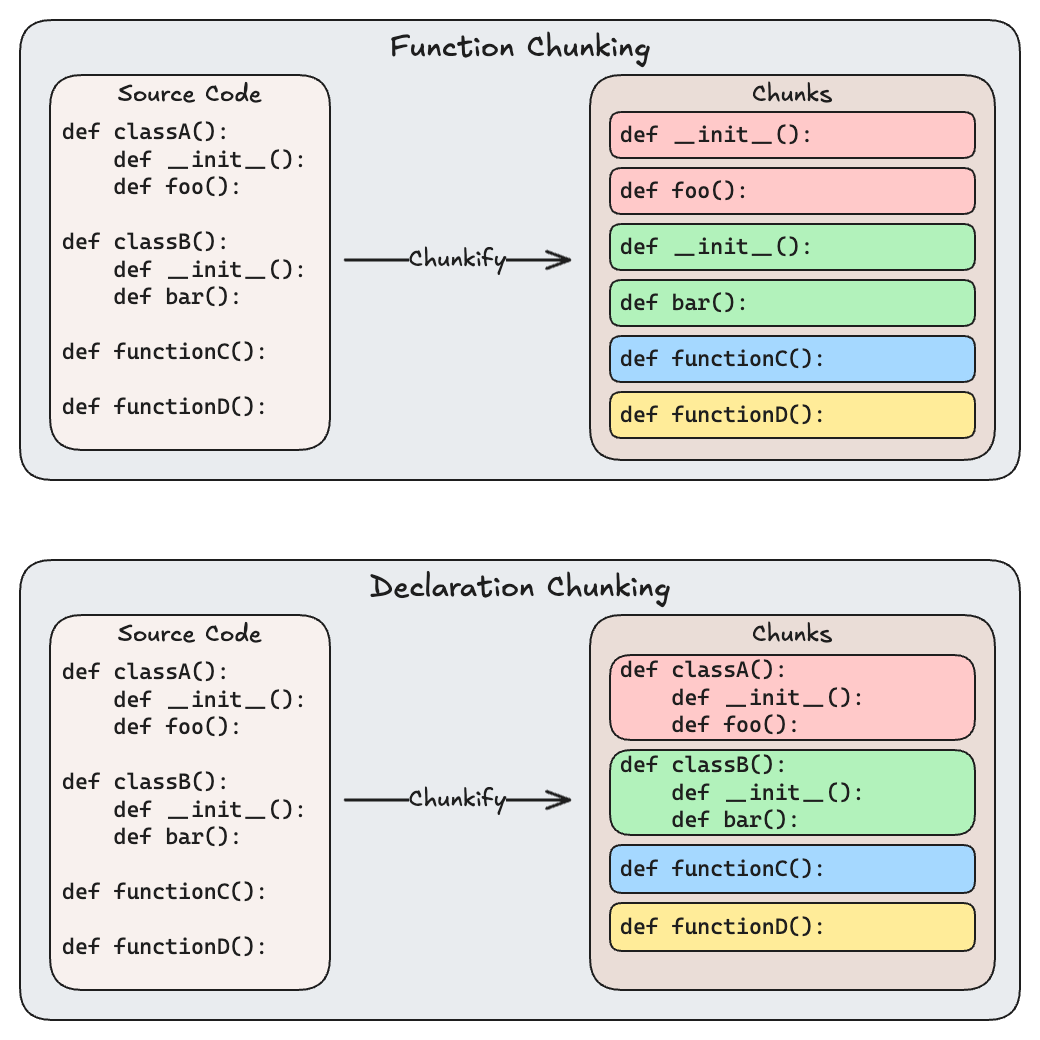}
  \caption{Function chunking vs.\ declaration chunking on the same Python source file. Function chunking extracts each function definition as a complete chunk (left), while declaration chunking retains only class headers, field definitions, and method signatures, omitting method bodies (right).}
  \label{fig:chunking-comparison}
\end{figure}

For source code, chunking presents additional challenges. Code has explicit syntactic structure (functions, classes, and modules) that provides natural segmentation boundaries, but these constructs vary in size and nesting depth across programming languages. Existing code chunking strategies range from structure-aware to structure-agnostic.
\gunel{I think this subsection will benefit from the examples for each type of chunking. I also find the description for function and declaration chunking not very easy to follow.}%
We describe four representative methods below and illustrate two of them (Function and Declaration) on the same source file in Figure~\ref{fig:chunking-comparison}. All Abstract Syntax Tree (AST)-based strategies are implemented via Tree-sitter\footnote{\url{https://github.com/tree-sitter/tree-sitter}}.

\subsubsection{Function Chunking}
\label{sec:function}

Function chunking traverses the AST to extract each top-level function or method definition as a single chunk, including both the signature and the complete body (Figure~\ref{fig:chunking-comparison}, left). \gunel{it is not very clear what ``extracted node'' is in this context.} When a function's text exceeds the maximum chunk size, it is split at its direct child-statement boundaries in source order: child statements are appended to the current chunk until the size budget is reached, then a new chunk begins. Function-level granularity is the default in CodeSearchNet~\cite{husainCodeSearchNetChallengeEvaluating2020} and is also adopted by DARE~\cite{sunDAREAligningLLM2026} for building retrieval corpora.

\subsubsection{Declaration Chunking}
\label{sec:declaration}

Rather than extracting full function bodies, declaration chunking targets class-level and function-level declaration nodes: each class node becomes a chunk containing the class header, field definitions, and method signatures, but not method bodies. Private functions (those whose names begin with an underscore in Python) are filtered out, retaining only the public interface of each module; this distinguishes declaration chunking from cAST, which applies no visibility-based filtering. Oversized declarations are split recursively using the same child-node procedure as function chunking. LongCodeZip~\cite{shiLongCodeZipCompressLong2025} employs a similar declaration-level granularity for context compression in code language models.

\subsubsection{Sliding Window Chunking}
\label{sec:sliding}
Sliding window chunking is the structure-agnostic baseline. It segments source files into fixed-size windows with configurable line overlap. The window advances by \emph{window\_length $-$ overlap\_lines} lines per step, and chunk boundaries align to line boundaries. This strategy requires no parsing and is language-independent, but can split functions mid-body or merge unrelated code fragments into the same chunk. Sliding window is the default chunking method in RepoCoder~\cite{zhangRepoCoderRepositoryLevelCode2023} and subsequent retrieval-augmented completion work.

\subsubsection{cAST Chunking}
\label{sec:cast}
cAST~\cite{zhangCASTEnhancingCode2025} is an AST-based chunking method that recursively splits large AST nodes and greedily merges adjacent sibling nodes within a size budget. Unlike function and declaration chunking, which target specific node types (functions, classes), cAST operates on all AST node types and optimizes for uniform chunk size. It uses its own assignment algorithm from the astchunk library, independent of the recursive splitting procedure shared by function and declaration chunking.

\subsection{Code Retrieval}
\label{sec:retrieval}

Code retrieval is the task of finding relevant code fragments from a corpus given a query. Formally, let $C = \{c_1, c_2, \dots, c_m\}$ denote a corpus of code fragments and $q$ a query. A retrieval function $\mathcal{R}$ returns the top-$k$ fragments ranked by a relevance score $s(q, c_i)$:
\begin{equation}
  \mathcal{R}(q, C, k) = \operatorname{top\text{-}k}_{c_i \in C} \; s(q, c_i)
\end{equation}

Retrieval methods differ in how they compute $s(q, c_i)$. \emph{Sparse} methods score relevance via weighted lexical overlap between tokenized query and document terms, requiring no training. \emph{Dense} methods encode $q$ and $c_i$ into continuous vectors via an embedding model and compute relevance as cosine similarity, capturing semantic relationships beyond token overlap.

Code retrieval has been a sustained research area. CodeSearchNet~\cite{husainCodeSearchNetChallengeEvaluating2020} established function-level code as the canonical retrieval unit; more recent suites such as CoIR~\cite{li_coir_nodate} and MTEB~\cite{muennighoff_mteb_2023} have begun to diversify beyond function-level granularity.

\subsection{Retrieval-Augmented Code Completion}
\label{sec:racc}

Repository-level code completion is the task of generating a code continuation at a cursor position within a file, given the broader context of the surrounding repository. Formally, given a repository $R = \{f_1, f_2, \dots, f_n\}$ and a target file $f_t$ with prefix $p$ (code before the cursor) and suffix $s$ (code after the cursor), the goal is to generate a continuation $y$ that correctly completes the code at the cursor position. In a retrieval-augmented setting, the system composes the chunking and retrieval stages into a three-stage pipeline:
\begin{equation}
  y = \mathcal{G}\!\bigl(p, s, \;\mathcal{R}(\,q,\;\textstyle\bigcup_{f \in R \setminus f_t} \mathcal{C}(f),\;k)\bigr)
\end{equation}
where $q$ is the query derived from the in-file context, $\mathcal{C}$ chunks cross-file source code into retrievable units, $\mathcal{R}$ retrieves the top-$k$ relevant chunks, and $\mathcal{G}$ denotes the code language model that generates $y$ conditioned on both the in-file context and the retrieved cross-file context.

The chunk-based pipeline is the most widely adopted approach to retrieval-augmented code completion~\cite{taoRetrievalAugmentedCodeGeneration2025}. Existing work primarily optimizes the retrieval or generation stages. RepoCoder~\cite{zhangRepoCoderRepositoryLevelCode2023} introduces an iterative retrieval-generation loop that refines the query using previously generated candidates. RLCoder~\cite{wang_rlcoder_2024} and AlignCoder~\cite{jiang_aligncoder_2026} train the retriever via reinforcement learning to select more useful cross-file context. Repoformer~\cite{wu_repoformer_2024} adds a selective retrieval mechanism that decides whether retrieval is beneficial before invoking it.

A separate line of work constructs explicit code graphs to capture cross-file dependencies. GraphCoder~\cite{liu_graphcoder_2024} builds a code context graph from control-flow and data-dependence relations, DRACO~\cite{cheng_dataflow-guided_2024} establishes a repository-specific context graph through dataflow analysis, and RepoGraph~\cite{ouyang_repograph_2025} maintains a repository-level structure graph for code intelligence tasks. However, building and maintaining such graphs requires substantial static analysis infrastructure, which is often impractical in realistic settings~\cite{taoRetrievalAugmentedCodeGeneration2025}. Our study therefore adopts chunk-based retrieval, which enables controlled comparison across chunking strategies without graph construction.

Two prior studies most closely relate to ours. cAST~\cite{zhangCASTEnhancingCode2025} proposes a new chunking method and evaluates it against a single baseline (Sliding Window) with one retriever and one generator, ablating chunk size but not cross-file context length. Galimzyanov et al.~\cite{galimzyanovPracticalCodeRAG2025} compare two strategies (Sliding Window and syntax-aware recursive splitting) across multiple retrievers and parameter configurations but fix the generator, leaving the interaction between chunking and generator choice unexamined. Our study differs from both by crossing a broader set of strategies (four) with all three pipeline dimensions jointly varied (four retrievers $\times$ five generators $\times$ nine parameter configurations), enabling us to isolate chunking's independent effect while controlling for retriever and generator variation.

\section{Research Questions}
\label{sec:rqs}

Existing retrieval-augmented code completion work optimizes the retriever or generator while treating chunking as a fixed preprocessing step~\cite{zhangRepoCoderRepositoryLevelCode2023, wang_rlcoder_2024, wu_repoformer_2024}. To isolate chunking's independent contribution, we control for retriever, generator, and parameter variation through a full-factorial design, forming a controlled empirical study around four research questions.

\textbf{RQ1 (Strategy Effect): How do different chunking strategies, both structure-aware and structure-agnostic, affect retrieval-augmented code completion quality?} This question establishes whether the choice of chunking strategy has a measurable effect on downstream Exact Match.

\textbf{RQ2 (Interaction Effect): How do chunking strategies interact with different retrieval methods (sparse vs.\ dense) and code completion models?} This question examines whether the chunking effect holds across retriever and generator combinations or depends on the choice of retriever and generator.

\textbf{RQ3 (Parameter Sensitivity): How sensitive is completion quality to chunking parameter settings (chunk size, overlap, cross-file context length), and what configurations are empirically justified?} This question ablates key parameters within each strategy to identify empirically justified defaults.

\textbf{RQ4 (Cost--Quality Trade-off): What is the token cost associated with different chunking strategies, and how does the cost--quality trade-off compare across configurations?} This question quantifies the practical cost of each strategy in tokens consumed, so that practitioners can balance quality against computational budget.

\section{Experiment Setup}
\label{sec:setup}

We construct a full-factorial experiment that crosses four chunking strategies with four retrievers (Section~\ref{sec:retrievers}), five code completion models (Section~\ref{sec:llms}), and nine parameter settings (Table~\ref{tab:parameters}) on two established benchmarks (Section~\ref{sec:benchmarks}). Because we restrict each benchmark to generators whose technical reports provide RAG-based results on that benchmark (Section~\ref{sec:llms}), this yields $4 \times 4 \times 4 \times 9 = 576$ settings on RepoEval and $4 \times 4 \times 2 \times 9 = 288$ on CrossCodeEval, totaling 864. A no-retrieval condition is included as a baseline in RQ1 to quantify the contribution of retrieval.

\subsection{Benchmarks}
\label{sec:benchmarks}

We evaluate on two established repository-level code completion benchmarks that require cross-file context.

\textbf{RepoEval}~\cite{zhangRepoCoderRepositoryLevelCode2023} provides completion tasks derived from recent commits across open-source Python repositories. Each task supplies a cursor position within a file and expects a single-line or multi-line completion that matches the ground truth. The benchmark provides line-level and API-level completion tasks; the latter specifically targets cross-file API invocations. \gunel{I find the next sentence hard to understand, what is the Python split? is it the tasks in the Python language?} The benchmark contains 1{,}600 line-level and 1{,}600 API-level instances (3{,}200 total).

\textbf{CrossCodeEval}~\cite{ding_crosscodeeval_2023} is a cross-file code completion benchmark constructed from multi-language repositories. \gunel{I think the "gold cross-file context" in the next sentence would benefit from more explanation.} Each instance is annotated with gold cross-file context, i.e., reference code fragments from other repository files that are relevant to the completion target. We use the Python subset, excluding instances whose source repositories or commits are no longer publicly accessible, yielding 1{,}937 of the original 2{,}665 instances. Since our study varies the chunking strategy, the pre-supplied cross-file context does not align with our chunk boundaries; we therefore re-chunk each repository and retrieve cross-file context from scratch using our own pipeline.

RepoEval serves as our primary evaluation benchmark; all main results (Sections~\ref{sec:rq1}--\ref{sec:rq4}) are reported on RepoEval. CrossCodeEval serves as a secondary benchmark to validate that the observed trends generalize beyond a single dataset.

\subsection{Retrievers}
\label{sec:retrievers}

We select one sparse and three dense retrievers \gunel{it is a spectrum or a binary choice?} to cover the two dominant retrieval paradigms---lexical matching (sparse) and semantic similarity (dense)---and test whether chunking effects are consistent across them. Table~\ref{tab:retrievers} lists the four retrievers and the no-retrieval baseline. \emph{No retrieval} serves as a baseline, providing the generator with only the in-file prefix and suffix (all code before and after the cursor in the target file, with no cross-file context) as supplied by the benchmarks \gunel{what does this context window include?}. We implement sparse retrieval using BM25~\cite{robertson_probabilistic_2009} via the BM25s library~\cite{luBM25SOrdersMagnitude2024}. For dense retrieval, we select three embedding models based on their ranking on the MTEB~\cite{muennighoff_mteb_2023} code-domain leaderboard: Qwen3-Embedding-0.6B and Qwen3-Embedding-4B~\cite{zhang_qwen3_2025} form a size-controlled pair within the same model family, and EmbeddingGemma-300M~\cite{vera_embeddinggemma_2025} provides a cross-family comparison point.

\begin{table}[h]
  \caption{Retrieval methods used in the experiment.}
  \label{tab:retrievers}
  \begin{tabular}{llr}
    \toprule
    Type & Model & Parameters \\
    \midrule
    None & --- & --- \\
    Sparse & BM25 & --- \\
    Dense & EmbeddingGemma-300M & 0.3B \\
    Dense & Qwen3-Embedding-0.6B & 0.6B \\
    Dense & Qwen3-Embedding-4B & 4B \\
    \bottomrule
  \end{tabular}
\end{table}

\subsection{Code Completion Models}
\label{sec:llms}
\jie{why not call the title Models?}
We select code completion models in the 6--9B parameter range: four code-specialized models and one general-purpose base model (Qwen3.5-9B). Constraining the parameter range holds model capacity approximately constant, so that observed differences across generators reflect architectural and training-data variation rather than scale effects. For each benchmark, we include only generators whose technical reports provide RAG-based evaluation results on that benchmark; results under comparable configurations serve as a sanity check on our implementation fidelity. This yields five generators in total: four for RepoEval and two for CrossCodeEval, with DeepSeek-Coder-6.7B~\cite{guoDeepSeekCoderWhenLarge2024} included in both (Table~\ref{tab:generators}). All models are served via vLLM~\cite{kwon_efficient_2023}.

\begin{table}[h]
  \caption{Code completion models used in the experiment.}
  \label{tab:generators}
  \begin{tabular}{lll}
    \toprule
    Model & Release Date & Benchmark \\
    \midrule
    DeepSeek-Coder-6.7B & Nov 2023 & Both \\
    StarCoder2-7B & Feb 2024 & CrossCodeEval \\
    Qwen2.5-Coder-7B & Sep 2024 & RepoEval \\
    Seed-Coder-8B & May 2025 & RepoEval \\
    Qwen3.5-9B & Feb 2026 & RepoEval \\
    \bottomrule
  \end{tabular}
\end{table}

\subsection{Configurations}
\label{sec:parameters}

Table~\ref{tab:parameters} summarizes the parameter space. We vary chunk size, the most direct parameter of a chunking strategy, over 1{,}000, 2{,}000, and 3{,}000 non-whitespace characters. We vary cross-file context length, which caps the total retrieved context prepended to the prompt, over 2{,}048, 4{,}096, and 8{,}192 tokens; smaller chunks allow more units within the same token budget, making the two dimensions interact. The full crossing yields nine parameter configurations per strategy--retriever--generator triple. All other parameters are fixed: maximum sequence and output lengths follow the technical reports of DeepSeek-Coder~\cite{guoDeepSeekCoderWhenLarge2024}, Qwen2.5-Coder~\cite{huiQwen25CoderTechnicalReport2024}, and Seed-Coder~\cite{seedSeedCoderLetCode2025}; overlap is fixed at 15 lines for Sliding Window; top-$k$ is fixed at 10; and temperature is set to 0 for reproducibility.
\begin{table}[t]
  \caption{Experimental parameter space. The two varied parameters are fully crossed ($3 \times 3 = 9$ configurations per strategy--retriever--generator triple); fixed parameters are held constant across all runs.}
  \label{tab:parameters}
  \begin{tabular}{lrl}
    \toprule
    Parameter & Values & Unit \\
    \midrule
    \multicolumn{3}{l}{\textit{Varied}} \\
    \quad Chunk size & 1{,}000 / 2{,}000 / 3{,}000 & nw-chars\textsuperscript{$\dagger$} \\
    \quad Cross-file context & 2{,}048 / 4{,}096 / 8{,}192 & tokens \\
    \midrule
    \multicolumn{3}{l}{\textit{Fixed}} \\
    \quad Sequence length & 8{,}192 & tokens \\
    \quad Output length & 50 & tokens \\
    \quad Overlap lines & 15 & lines \\
    \quad Top-$k$ & 10 & \\
    \quad Temperature & 0 & \\
    \bottomrule
    \multicolumn{3}{l}{\textsuperscript{$\dagger$}\small Non-whitespace characters, following cAST~\cite{zhangCASTEnhancingCode2025}.}
  \end{tabular}
\end{table}

\subsection{Evaluation Metric}
\label{sec:metric}

We report Exact Match (EM) \jie{sorry this just came to my thought: why not use llm as a judge?}as the primary evaluation metric. EM awards a score of 1 when the generated completion is identical to the ground truth after whitespace normalization, and 0 otherwise. We adopt EM because it is the standard metric for retrieval-augmented code completion: both benchmarks~\cite{zhangRepoCoderRepositoryLevelCode2023, ding_crosscodeeval_2023} and all prior systems we compare against~\cite{wang_rlcoder_2024, wu_repoformer_2024, zhangCASTEnhancingCode2025, guoDeepSeekCoderWhenLarge2024, huiQwen25CoderTechnicalReport2024} report EM, ensuring direct comparability. As a strict metric, \jie{find any proof to show how widely this metric is adopted to back up our choice. Otherwise it is very risky}it gives no partial credit for semantically correct but surface-different completions; we discuss the implications of this choice in Section~\ref{sec:threats}.

\section{Results}
\label{sec:results}

We present results organized by research question. RepoEval is the primary benchmark; CrossCodeEval results follow each subsection as a secondary validation.

\subsection{RQ1: Strategy Effect}
\label{sec:rq1}
\jie{at the beginning of each RQ, briefly remind readers what this RQ is about, such as ``RQ1 studies...''}
RQ1 measures the main effect of chunking strategy on code completion quality, averaging over all retriever, generator, and parameter combinations.

\begin{table}[t]
  \caption{Mean Exact Match (\%) per chunking strategy, averaged across all retriever--generator--parameter configurations (144 for RepoEval, 72 for CCEval). Best result per column is \textbf{bold}.}
  \label{tab:rq1}
  \begin{tabular}{l rrr}
    \toprule
    Strategy & RepoEval API & RepoEval Line & CCEval\textsuperscript{$\dagger$} \\
    \midrule
    No Retrieval & 34.13 & 43.80 & 9.14 \\
    Function & 42.27 & 51.27 & 24.21 \\
    Declaration & 45.85 & 54.84 & 27.71 \\
    cAST & 45.93 & 56.54 & 28.19 \\
    Sliding Window & \textbf{46.23} & \textbf{56.91} & \textbf{28.40} \\
    \bottomrule
    \multicolumn{4}{l}{\textsuperscript{$\dagger$}\small CCEval = CrossCodeEval.}
  \end{tabular}
\end{table}

Table~\ref{tab:rq1} reports the mean Exact Match for each chunking strategy, where each cell averages over all retriever, generator, and parameter-setting combinations: $4 \times 4 \times 9 = 144$ configurations for RepoEval and $4 \times 2 \times 9 = 72$ for CCEval.

\paragraph{RepoEval.} All four retrieval-augmented strategies outperform the no-retrieval baseline by 8.14--12.10\,pp on API-level and 7.47--13.11\,pp on line-level completion. Function chunking ranks last on both splits, trailing Sliding Window by 3.96\,pp on API-level and 5.64\,pp on line-level. Declaration, cAST, and Sliding Window cluster within 0.38\,pp on API-level and 2.07\,pp on line-level, with Sliding Window leading narrowly. Pairwise Wilcoxon signed-rank tests with Bonferroni correction confirm that all six pairwise differences are statistically significant ($p < 0.05$) on both splits; Function versus any other strategy yields Cliff's $\delta = -1.0$ (large effect). Function's underperformance stems from two factors: individual functions are typically shorter than the smallest chunk-size setting, underutilizing the retrieval budget (RQ3 confirms Function is insensitive to chunk size, within 0.6\,pp), and Function chunking discards module-level code outside function bodies that other strategies retain.

\paragraph{CCEval.} CCEval replicates the same ranking: Sliding Window leads at 28.40\%, followed by cAST (28.19\%), Declaration (27.71\%), and Function (24.21\%). All four strategies improve substantially over the no-retrieval baseline (9.14\%), with gains of 15.07--19.26\,pp. The gap between Function and the remaining strategies (3.50--4.19\,pp) mirrors the RepoEval pattern. \gunel{It seems CCEval results are much worse than RepoEval ones. I think we need to comment on this in some way.}The absolute EM scores are lower than on RepoEval because CCEval was constructed from post-March-2023 repositories with explicit training-data leakage filtering, and three of five generators achieve near-zero no-retrieval baselines on CCEval (Section~\ref{sec:cceval_limitations}).

\begin{tcolorbox}[colback=blue!10, colframe=black, arc=4pt, boxrule=0.5pt]
\textbf{RQ1:} \textit{Chunking strategy has a statistically significant and practically meaningful effect on code completion quality. Function chunking underperforms all other strategies by 3.57--5.64\,pp EM on RepoEval (Cliff's $\delta=-1.0$). Declaration, Sliding Window, and cAST perform comparably, with Sliding Window holding a slight but consistent lead across both benchmarks (all three evaluation splits).}
\end{tcolorbox}

\subsection{RQ2: Interaction Effect}
\label{sec:rq2}

\gunel{I think it needs to be explained slightly better that Tables 5 and 6 are for RepoEval, and tables 7 and 8 are for CCEval.} RQ2 examines whether the strategy ranking from RQ1 is stable across different retrievers and generators. Tables~\ref{tab:rq2a}--\ref{tab:rq2b} disaggregate the RQ1 averages by retriever and by generator on RepoEval; Tables~\ref{tab:rq2a_cceval}--\ref{tab:rq2b_cceval} present the corresponding CCEval breakdowns.

\begin{table}[t]
  \caption{Mean Exact Match (\%) per chunking strategy and retriever on RepoEval, averaged across all generator and parameter configurations (36 per cell). \jie{turned following into foot note to save space? for all tables}Best result per column is \textbf{bold}.}
  \label{tab:rq2a}
  \begin{tabular}{l rrrr}
    \toprule
    Strategy & BM25 & EmbGem. & Qwen-0.6B & Qwen-4B \\
    \midrule
    \multicolumn{5}{c}{\textit{API-level}} \\
    Function & 42.53 & 42.37 & 42.41 & 41.76 \\
    Declaration & 46.33 & 45.99 & 45.47 & 45.60 \\
    cAST & 46.61 & 46.08 & 45.52 & 45.54 \\
    Sliding Window & \textbf{46.74} & \textbf{46.20} & \textbf{45.84} & \textbf{46.13} \\
    \midrule
    \multicolumn{5}{c}{\textit{Line-level}} \\
    Function & 51.11 & 51.36 & 51.32 & 51.29 \\
    Declaration & 54.72 & 55.08 & 54.74 & 54.81 \\
    cAST & 56.87 & 56.48 & 56.48 & 56.33 \\
    Sliding Window & \textbf{57.62} & \textbf{56.66} & \textbf{56.51} & \textbf{56.84} \\
    \bottomrule
    \multicolumn{5}{l}{\textsuperscript{$\ddagger$}\footnotesize EmbGem.=EmbeddingGemma; Qwen-*=Qwen3-Embedding.}
  \end{tabular}
\end{table}

\begin{table}[t]
  \caption{Mean Exact Match (\%) per chunking strategy and generator on RepoEval, averaged across all retriever and parameter configurations (36 per cell). Best result per column is \textbf{bold}.}
  \label{tab:rq2b}
  \setlength{\tabcolsep}{3pt}
  \begin{tabular}{l rrrr}
    \toprule
    Strategy & DSCoder & Qwen2.5 & Qwen3.5 & SeedCoder \\
    \midrule
    \multicolumn{5}{c}{\textit{API-level}} \\
    Function & 39.65 & 43.31 & 42.14 & 43.98 \\
    Declaration & 43.55 & 47.15 & 45.67 & 47.02 \\
    cAST & 43.49 & 47.34 & 45.82 & 47.09 \\
    Sliding Window & \textbf{44.18} & \textbf{47.50} & \textbf{46.14} & \textbf{47.10} \\
    \midrule
    \multicolumn{5}{c}{\textit{Line-level}} \\
    Function & 48.58 & 52.30 & 50.22 & 53.99 \\
    Declaration & 52.79 & 55.20 & 54.21 & 57.15 \\
    cAST & 54.64 & 56.92 & 56.19 & 58.40 \\
    Sliding Window & \textbf{55.17} & \textbf{57.12} & \textbf{56.44} & \textbf{58.90} \\
    \bottomrule
  \end{tabular}
\end{table}

\begin{table}[t]
  \caption{Mean Exact Match (\%) per chunking strategy and retriever on CCEval\textsuperscript{$\dagger$}, averaged across all generator and parameter configurations (18 per cell). Column abbreviations follow Table~\ref{tab:rq2a}. Best result per column is \textbf{bold}.}
  \label{tab:rq2a_cceval}
  \begin{tabular}{l rrrr}
    \toprule
    Strategy & BM25 & EmbGem. & Qwen-0.6B & Qwen-4B \\
    \midrule
    Function & 23.40 & 23.98 & 25.10 & 24.35 \\
    Declaration & 26.53 & 27.44 & 28.71 & 28.14 \\
    cAST & 26.84 & \textbf{28.42} & 28.73 & \textbf{28.79} \\
    Sliding Window & \textbf{27.60} & 28.23 & \textbf{29.15} & 28.60 \\
    \bottomrule
    \multicolumn{5}{l}{\textsuperscript{$\dagger$}\small CCEval = CrossCodeEval.}
  \end{tabular}
\end{table}

\begin{table}[t]
  \caption{Mean Exact Match (\%) per chunking strategy and generator on CCEval, averaged across all retriever and parameter configurations (36 per cell). Best result per column is \textbf{bold}.}
  \label{tab:rq2b_cceval}
  \begin{tabular}{l rr}
    \toprule
    Strategy & DSCoder & StarCoder2 \\
    \midrule
    Function & 25.63 & 22.78 \\
    Declaration & 29.01 & 26.40 \\
    cAST & 29.64 & 26.74 \\
    Sliding Window & \textbf{29.91} & \textbf{26.88} \\
    \bottomrule
  \end{tabular}
\end{table}

\jie{no need to use seperate paras for two datasets. No need to highlight every number in the table}The strategy ranking holds across all four retrievers and all four generators on both benchmarks (Tables~\ref{tab:rq2a}--\ref{tab:rq2b} for RepoEval; Tables~\ref{tab:rq2a_cceval}--\ref{tab:rq2b_cceval} for CCEval). On RepoEval, switching retrievers within a strategy changes EM by at most 1.11\,pp, far less than the 3.43--6.51\,pp gap between the best and worst strategy within any single retriever. Generators introduce larger absolute variation (up to 5.41\,pp), but the ranking is unchanged: Function is last and Sliding Window is first across all retriever--split and generator--split combinations. CCEval replicates both patterns: within-strategy retriever variation reaches 2.18\,pp, while within-retriever strategy variation ranges from 4.05 to 4.44\,pp. Sliding Window and cAST trade the lead across retrievers (within 0.19\,pp), but Function consistently ranks last under both generators.

\begin{tcolorbox}[colback=blue!10, colframe=black, arc=4pt, boxrule=0.5pt]
\textbf{RQ2:} \textit{The chunking strategy ranking is stable across all retriever and generator combinations. Retriever choice accounts for $\leq$1.11\,pp EM variation on RepoEval, far less than the 3.43--6.51\,pp gap between strategies. Generators shift absolute EM more than retrievers, but do not alter the strategy ranking.}
\end{tcolorbox}

\subsection{RQ3: Parameter Sensitivity}
\label{sec:rq3}

\begin{figure*}[t]
  \centering
  \includegraphics[width=\textwidth]{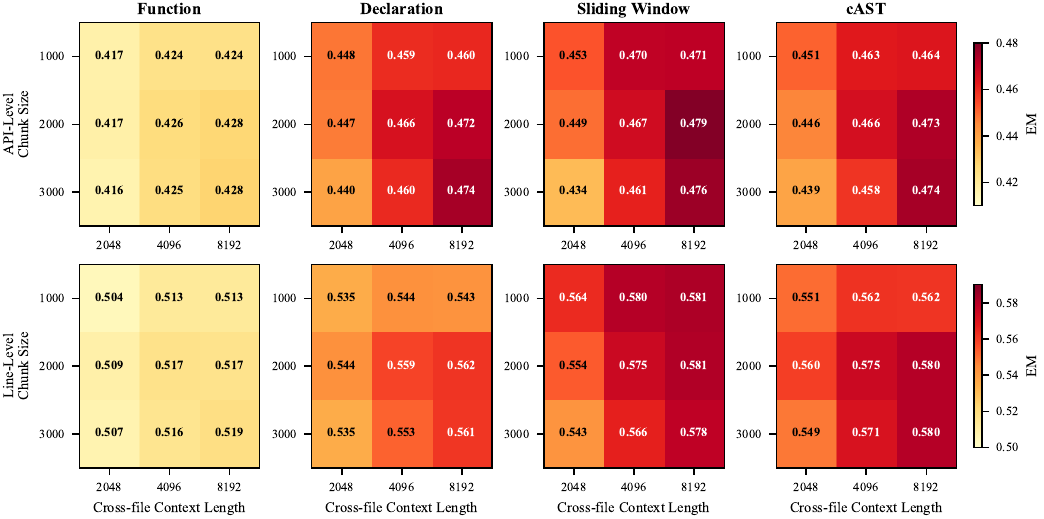}
  \caption{Exact Match on RepoEval as a function of cross-file context length in tokens (x-axis) and chunk size in non-whitespace characters (y-axis), for each chunking strategy (columns) and completion level (rows). Each cell averages over all retriever--generator combinations (16 per cell). Color encodes EM; darker is higher.}
  \label{fig:rq3_heatmap}
\end{figure*}

\begin{figure}[t]
  \centering
  \includegraphics[width=\columnwidth]{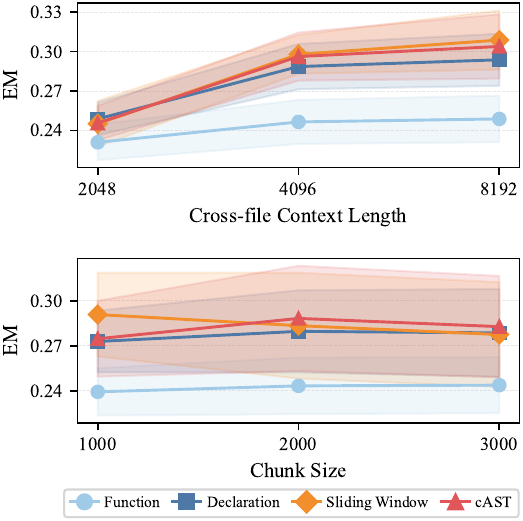}
  \caption{Exact Match on CCEval as a function of cross-file context length in tokens (top) and chunk size in non-whitespace characters (bottom), averaged across all retriever--generator combinations. Shaded regions denote $\pm$1 standard deviation.}
  \label{fig:rq3_line}
\end{figure}

RQ3 investigates the sensitivity of completion quality to chunk size and cross-file context length. Figure~\ref{fig:rq3_heatmap} shows EM on RepoEval as a heat map over chunk sizes (1{,}000, 2{,}000, 3{,}000 non-whitespace characters) and cross-file context lengths (2{,}048, 4{,}096, 8{,}192 tokens). Figure~\ref{fig:rq3_line} presents the same parameters on CCEval.

\paragraph{Cross-file context.} On RepoEval, increasing context length from 2{,}048 to 8{,}192 tokens monotonically improves EM for all four strategies, with gains from 0.7\,pp (Function) to 4.2\,pp (Sliding Window). The step from 2{,}048 to 4{,}096 contributes more than the further step to 8{,}192, indicating diminishing returns. Pairwise Wilcoxon signed-rank tests with Bonferroni correction confirm all three context-length differences are statistically significant ($p < 0.05$) with medium-to-large effect sizes ($\delta = 0.46$--$0.99$) on both splits. CCEval exhibits the same monotonic trend (Figure~\ref{fig:rq3_line}): Declaration, Sliding Window, and cAST rise from approximately 25\% to 30\%, while Function increases by only 1\,pp, confirming that Function chunking cannot fully exploit larger context budgets.

\paragraph{Chunk size.} The effect of chunk size depends on context length. At 2{,}048 tokens, larger chunks reduce EM by up to 1.9\,pp because fewer distinct chunks fit the budget; at 8{,}192 tokens this penalty disappears. The effect is non-monotonic: EM peaks at chunk size 2{,}000 and declines at 3{,}000 ($\delta = -0.38$ to $-0.43$). Function chunking is insensitive to chunk size (within 0.6\,pp) because individual functions are typically shorter than the smallest setting. On CCEval, chunk size has a negligible effect ($\leq$1\,pp). Across both benchmarks, 2{,}000 non-whitespace characters is a robust default.

\begin{figure}[t]
  \centering
  \includegraphics[width=\columnwidth]{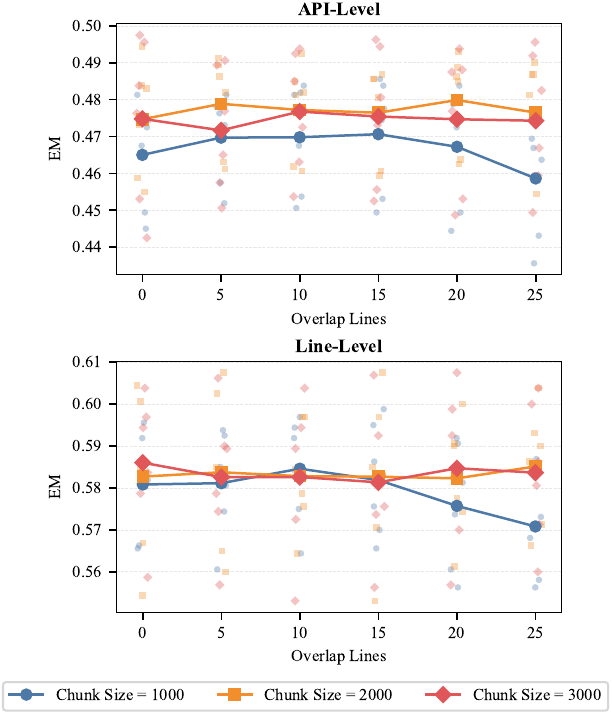}
  \caption{Effect of sliding window overlap (0--25 lines) on Exact Match for RepoEval API-level (top) and line-level (bottom) completion, at three chunk sizes. Lines connect the mean EM across 6 retriever--generator combinations; scattered points show individual combinations. Cross-file context length is fixed at 8{,}192 tokens.}
  \label{fig:ablation_overlap}
\end{figure}

\jie{why call them ablation? Cannot we turn them into RQs or merge with existing RQs?}
\paragraph{Overlap (Sliding Window only).} Sliding Window is the only strategy with an overlap parameter. Figure~\ref{fig:ablation_overlap} shows the effect of varying overlap from 0 to 25 lines in increments of 5 on RepoEval, with context length fixed at 8{,}192 tokens. Since RQ2 established that retriever and generator choice have limited effect on the strategy ranking, we restrict this analysis to 6 representative retriever--generator combinations. For chunk sizes 2{,}000 and 3{,}000, overlap has a negligible effect: EM varies by less than 0.5\,pp across the entire 0--25 range on both completion levels. For chunk size 1{,}000, moderate overlap (5--15 lines) yields up to 0.5\,pp improvement on API-level, but 25-line overlap degrades EM by 1.2\,pp because large overlap reduces effective new content per chunk and retrieval diversity within the fixed context budget. The 15-line default used in the main experiments falls within the stable plateau for all chunk sizes, confirming that this choice does not bias the main results.

\begin{tcolorbox}[colback=blue!10, colframe=black, arc=4pt, boxrule=0.5pt]
\textbf{RQ3:} \textit{Cross-file context length is the dominant parameter (up to 4.2\,pp EM, $\delta \geq 0.46$), with diminishing returns beyond 4{,}096 tokens. Chunk size has a weaker, non-monotonic effect ($\leq$1.9\,pp); overlap is negligible ($\leq$0.5\,pp) for chunk sizes $\geq$2{,}000. All trends confirmed on CCEval. Recommended default: chunk size 2{,}000, context length $\geq$4{,}096.}
\end{tcolorbox}

\subsection{RQ4: Cost--Quality Trade-off}
\label{sec:rq4}

\begin{figure}[t]
  \centering
  \includegraphics[width=\columnwidth]{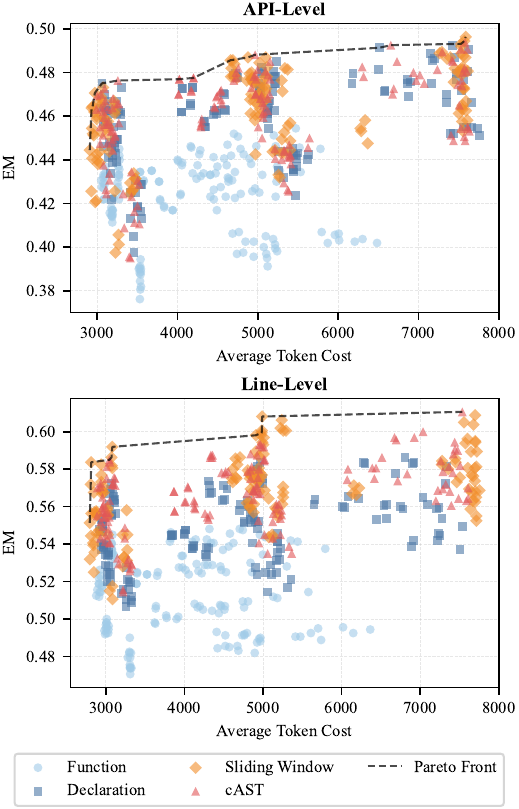}
  \caption{Exact Match vs.\ average token cost on RepoEval for API-level (top) and line-level (bottom) completion. Each point represents one configuration (strategy $\times$ retriever $\times$ generator $\times$ parameters). The dashed line traces the Pareto front.}
  \label{fig:rq4_repoeval}
\end{figure}

\begin{figure}[t]
  \centering
  \includegraphics[width=\columnwidth]{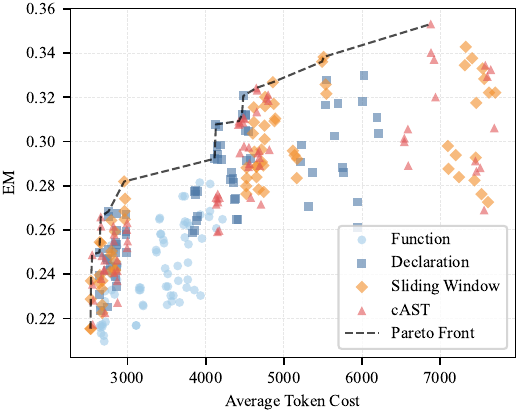}
  \caption{Exact Match vs.\ average token cost on CCEval. Each point represents one configuration. The dashed line traces the Pareto front.}
  \label{fig:rq4_cceval}
\end{figure}

RQ4 quantifies the token cost of each strategy and identifies Pareto-optimal configurations. Figures~\ref{fig:rq4_repoeval} and~\ref{fig:rq4_cceval} plot all 576 RepoEval and 288 CCEval configurations in the cost--quality plane (average token cost per prompt vs.\ Exact Match).

\paragraph{RepoEval.} The Pareto front is dominated by Sliding Window and cAST configurations on both completion levels, with occasional Declaration points at intermediate budgets. Function chunking never appears on the Pareto front: its lower token cost does not compensate for the 3.96--5.64\,pp EM deficit (RQ1). The front rises steeply from approximately 3{,}000 to 5{,}000 tokens, then flattens, reflecting the diminishing returns observed in RQ3. Practitioners operating under a fixed token budget gain the most improvement by scaling from 2{,}048 to 4{,}096 tokens. The best Pareto-optimal configurations reach 49\% EM on API-level and 60\% on line-level at approximately 7{,}500 tokens, using Sliding Window or cAST with chunk size 2{,}000 and context length 8{,}192.

\paragraph{CCEval.} CCEval confirms the same pattern (Figure~\ref{fig:rq4_cceval}). The Pareto front is again composed of Sliding Window and cAST configurations, with Function consistently below the front. Declaration appears at low-budget points, where its compact chunks achieve competitive EM at reduced token cost. The same inflection around 5{,}000 tokens is visible: below this threshold the front rises steeply, above it returns diminish. At the highest budgets ($>$7{,}000), several configurations fall below the front, indicating that the largest context settings can introduce noise. The consistency across both benchmarks strengthens the RQ3 recommendation: practitioners should allocate at least 4{,}096 tokens of cross-file context using Sliding Window or cAST at chunk size 2{,}000.

\begin{tcolorbox}[colback=blue!10, colframe=black, arc=4pt, boxrule=0.5pt]
\textbf{RQ4:} \textit{Sliding Window and cAST dominate the Pareto front on both benchmarks, achieving the best EM at every token budget. Function chunking is never Pareto-optimal: its lower token cost does not offset its consistently lower quality. Despite diminishing returns beyond approximately 5{,}000 tokens, the Pareto front rises monotonically with context budget, indicating that practitioners who can afford the token cost should prefer larger cross-file context.}
\end{tcolorbox}

\section{Discussion}
\label{sec:discussion}

This section reports three additional analyses---cross-language validation (Section~\ref{sec:ablation_java}), partial-match patterns (Section~\ref{sec:ablation_partial}), and CrossCodeEval's methodological constraints (Section~\ref{sec:cceval_limitations})---followed by broader implications (Section~\ref{sec:implications}).

\subsection{Cross-Language Validation (Java)}
\label{sec:ablation_java}


All main experiments evaluate on Python only. To assess cross-language generalizability, we replicate the RQ1 and RQ3 analyses on the Java subset of CrossCodeEval (1{,}028 instances after the same accessibility filtering applied to the Python subset) using the same pipeline, retrievers, generators, and parameter configurations.

The main findings from Python replicate on Java. Function chunking remains the weakest strategy at 22.31\% mean EM, trailing the other three by 4.20--5.31\,pp. The monotonic benefit of longer cross-file context holds: all four strategies improve from 2{,}048 to 8{,}192 tokens, with gains of 1.5--4.8\,pp that mirror the Python range (0.7--4.2\,pp on RepoEval). Chunk size remains a negligible factor on Java, as on Python. One difference emerges in the ranking among the top three strategies: cAST (27.62\%) leads on Java, overtaking Declaration (26.94\%) and Sliding Window (26.51\%), whereas on the Python subset Sliding Window ranks first. The gap between cAST and Sliding Window (1.11\,pp) is larger than on Python (0.21\,pp in the opposite direction), suggesting that structure-aware chunking may benefit more from Java's explicit block structure and deeper nesting. Absolute EM on Java (22--28\%) is comparable to the Python subset of CCEval (24--28\%), indicating that the lower scores relative to RepoEval reflect the benchmark's stricter filtering rather than a language effect. Despite this shift, the top three strategies remain within 1.11\,pp of each other on Java, far closer than their shared gap over Function chunking. As on both Python benchmarks, Function chunking is never Pareto-optimal on Java: its lower token cost does not compensate for the 4.20--5.31\,pp EM deficit. The core finding therefore generalizes across languages: Function chunking should be avoided regardless of the target language, while the choice among the remaining strategies can be made on practical grounds (e.g., parsing requirements, language support) rather than quality.

\subsection{Partial-Match Patterns}
\label{sec:ablation_partial}


EM is binary, so this analysis examines whether chunking strategies differ in partial-match behavior when EM\,=\,0. For each such instance, we truncate the prediction to the ground-truth length and compute a character-level match indicator at every position, aggregated across 48 configuration groups (2 retrievers $\times$ 4 generators $\times$ 2 splits $\times$ 3 chunk sizes) at context length 8{,}192.

All four strategies exhibit the same prefix-dominant decay pattern: match rates start above 91\% at the first 5\% of the ground truth and decline monotonically to 7--9\% at the final 5\%. The strategies do not differ in \emph{where} they match the ground truth; they differ in \emph{how deep} the match extends. Sliding Window maintains the highest match rate at every position (92.9\% at 0--5\%, 9.4\% at 95--100\%), followed by cAST, Declaration, and Function (91.8\% at 0--5\%, 6.6\% at 95--100\%). The mean prefix match ratio, defined as the fraction of ground truth characters matched contiguously from the start, follows the same ranking: 0.464 for Sliding Window, 0.459 for cAST, 0.431 for Declaration, and 0.379 for Function. Sliding Window completions thus stay correct for 22\% more of the ground truth than Function completions before diverging.


The gap is amplified on the $n$\,=\,11{,}162 instances where methods disagree on EM: by position 50--55\%, Sliding Window maintains 60.4\% while Function drops to 38.4\%; at position 95--100\%, the difference reaches 28.7\,pp. This confirms that EM disagreements arise from differences in how far the generator sustains correct sequences, not from qualitatively different match patterns. The consistency of the strategy ranking between EM and prefix match ratio (Sliding Window $>$ cAST $>$ Declaration $>$ Function under both metrics) provides evidence that EM, despite being a strict binary metric, faithfully reflects the underlying quality ordering.

\subsection{CrossCodeEval as a Secondary Benchmark}
\label{sec:cceval_limitations}

CrossCodeEval is used as a secondary validation benchmark for two reasons. First, 728 of the original 2{,}665 instances (27\%) reference repositories or commits that are no longer publicly accessible, introducing a selection bias toward repositories with stable hosting.

Second, three of five generators (Qwen2.5-Coder-7B, Qwen3.5-9B, Seed-Coder-8B) achieve no-retrieval baseline EM below 3.1\% on CrossCodeEval (Table~\ref{tab:cceval_baseline}), raising a concern that retrieval-augmented gains could be confounded by a weak prior on the target completion style. We restrict the CrossCodeEval evaluation to the two generators with the strongest baselines (DeepSeek-Coder-6.7B at 10.43\% and StarCoder2-7B at 7.85\%) and verify that the strategy ranking remains consistent with RepoEval.

\begin{table}[h]
  \caption{No-retrieval baseline Exact Match (\%) on CrossCodeEval per generator. Models below the dashed line are excluded from CrossCodeEval experiments.}
  \label{tab:cceval_baseline}
  \begin{tabular}{l r}
    \toprule
    Generator & EM (\%) \\
    \midrule
    DeepSeek-Coder-6.7B & 10.43 \\
    StarCoder2-7B & 7.85 \\
    \midrule
    Qwen2.5-Coder-7B & 3.05 \\
    Seed-Coder-8B & 2.79 \\
    Qwen3.5-9B & 2.43 \\
    \bottomrule
  \end{tabular}
\end{table}

\subsection{Implications}
\label{sec:implications}

\paragraph{Function-level granularity in code retrieval benchmarks.} Function chunking consistently underperforms all other strategies by 3.57--5.64\,pp EM (RQ1) and is never Pareto-optimal (RQ4). This finding has implications beyond chunking configuration: many widely used code retrieval benchmarks, including CodeSearchNet~\cite{husainCodeSearchNetChallengeEvaluating2020}, adopt function-level granularity as both the retrieval corpus and the unit of relevance judgment. Our results suggest that this granularity may not be optimal for downstream code completion tasks, where finer-grained or mixed-granularity chunks yield higher completion EM\@. Recent code retrieval benchmarks such as CoIR~\cite{li_coir_nodate} and MTEB~\cite{muennighoff_mteb_2023} have begun to diversify corpus granularity beyond the function level, a direction our findings empirically support.

\paragraph{Code LLMs tolerate longer context despite diminishing relevance.} Increasing cross-file context length from 2{,}048 to 8{,}192 tokens monotonically improves EM by up to 4.2\,pp across all strategies (RQ3), even though a larger context budget fills the prompt with progressively lower-ranked chunks. Meanwhile, swapping the retriever shifts EM by at most 1.11\,pp (RQ2), regardless of whether the retriever is sparse BM25 or a dense embedding model. These two observations suggest that code LLMs can tolerate additional context of mixed relevance and still extract useful signal. We note, however, that we measure this effect only through end-to-end EM; we do not evaluate retrieval precision directly. Whether the generator genuinely attends to lower-ranked chunks or simply benefits from a higher probability of including at least one relevant fragment is an open question that warrants future investigation using attention-level analysis.

\paragraph{Chunking as context compression.} Our results show that recall-oriented concerns (strategy choice, retriever quality) have a smaller effect on downstream EM than the total volume of context provided to the generator (RQ2--RQ3). Concretely, swapping retrievers shifts EM by at most 2.18\,pp, while doubling the context budget yields up to 4.2\,pp. As context windows grow, chunking's role may shift from determining \textbf{what} to retrieve toward determining \textbf{how} to compress a token budget into the most informative context. Structure-aware strategies whose advantage as retrieval units is marginal over Sliding Window could prove more valuable as compression tools, selecting which parts of a file to retain and which to discard. Evaluating chunking as a compression mechanism remains an open direction for future work.

\section{Threats to Validity}
\label{sec:threats}

\paragraph{External validity.}
The main experiments evaluate on Python only. To mitigate this, we replicate the RQ1 and RQ3 analyses on the Java subset of CrossCodeEval (Section~\ref{sec:ablation_java}): the strategy ranking is consistent across both languages, though cAST and Sliding Window swap positions on Java. All generators fall in the 6--9B parameter range; larger models with longer effective context windows may interact differently with chunking strategies, and our findings should be validated at other scales.

\paragraph{Internal validity.}
All five generators were trained on corpora that may include RepoEval or CrossCodeEval repositories, potentially inflating absolute EM. However, contamination affects all strategies equally within a generator and does not alter relative rankings; the no-retrieval baseline captures any contamination-driven advantage. The consistency of rankings across five generators with different training corpora further reduces this concern.

We fix top-$k$ at $k{=}10$ following RepoCoder~\cite{zhangRepoCoderRepositoryLevelCode2023} and do not ablate this parameter. Function chunking leaves unused context budget at $k{=}10$ and 8{,}192 tokens, so a larger $k$ could improve its performance. However, the performance gap persists at 2{,}048 and 4{,}096 tokens where Function fills the budget (RQ3), indicating the gap is not driven by retrieval volume alone.

All generators use greedy decoding (temperature\,=\,0); we ran a subset of configurations multiple times and observed negligible variation, so we report single-run results.

\paragraph{Construct validity.}
Exact Match awards no partial credit for semantically correct but surface-different completions. This bias may particularly affect Function chunking, where EM could undercount functionally equivalent but syntactically different completions. Strategy rankings under Edit Similarity (ES), a partial-credit metric based on normalized edit distance, are consistent with EM across all strategies and benchmarks (detailed ES results in the replication package\footnote{\url{https://doi.org/10.5281/zenodo.19228777}}), and the partial-match analysis in Section~\ref{sec:ablation_partial} corroborates this: prefix match ratio ranks strategies in the same order as EM. Complementary metrics such as CodeBLEU~\cite{ren_codebleu_2020}, functional correctness (pass@$k$)~\cite{chen_evaluating_2021}, or LLM-as-a-judge evaluation could reveal quality differences that surface-level metrics miss; we leave these extensions to future work.

\section{Conclusion}

This paper presents the first controlled empirical study isolating the chunking dimension in retrieval-augmented code completion. Across 864 settings---576 on RepoEval and 288 on CrossCodeEval---spanning four chunking strategies, four retrievers, five generators, and nine parameter configurations, we find that chunking strategy has a statistically significant effect on Exact Match, with Function chunking underperforming by 3.57--5.64\,pp; that cross-file context length is the dominant tuning lever (up to 4.2\,pp gain), far exceeding the effect of retriever choice ($\leq$1.11\,pp on RepoEval); and that Sliding Window and cAST dominate the cost--quality Pareto front.

For practitioners, the primary recommendation is to maximize the cross-file context budget and avoid function-level chunking. Perhaps the most actionable finding is that structure-aware chunking (cAST, Declaration) does not outperform the simple, language-independent Sliding Window on either quality or cost efficiency. Sliding Window and cAST are both strong defaults: they perform within 1.11\,pp of each other across Python and Java and require no language-specific tuning beyond a chunk size of 2{,}000 non-whitespace characters.

Future work should validate these findings at larger model scales, where longer effective context windows may alter the interaction between chunking and generation, and with complementary metrics such as functional correctness (pass@$k$) that capture semantic equivalence beyond surface matching. As context windows continue to grow, evaluating chunking as a \emph{compression} mechanism that allocates a token budget for maximum informativeness may prove more consequential than its current role as a retrieval mechanism.


\bibliographystyle{ACM-Reference-Format}
\bibliography{Chunk}


\end{document}